\documentclass{article}

\usepackage{natbib}
\usepackage{url}            
\usepackage{amsfonts}       
\usepackage{nicefrac}       
\usepackage{graphicx}
\usepackage{doi}
\usepackage{subfigure}
\usepackage{amsthm}
\usepackage{bbm}
\usepackage{algorithm}
\usepackage{algpseudocode}
\usepackage{hyperref}
\usepackage[english]{babel}
\begin{document}
\title{A Probabilistic Approach to The Perfect Sum Problem}

\newcommand{\authorlastnames}{AuthorLastName}

\newcommand{\shortarticletitle}{A Probabilistic Approach to The Perfect Sum Problem}

\author{Kristof Pusztai\\
        \href{mailto:k.pusztai@columbia.edu}{k.pusztai@columbia.edu}\\
         Department of Statistics,  Columbia University\\
        1255 Amsterdam Ave, New York, NY 10027}

\date{{\small \today }}
\maketitle

\begin{abstract}
	The subset sum problem is known to be an NP-hard problem in the field of computer science with the fastest known approach having a run-time complexity of $O(2^{0.3113n})$. A modified version of this problem is known as the perfect sum problem and extends the subset sum idea further. This extension results in additional complexity, making it difficult to compute for a large input. In this paper, I propose a probabilistic approach which approximates the solution to the perfect sum problem by approximating the distribution of potential sums. Since this problem is an extension of the subset sum, our approximation also grants some probabilistic insight into the solution for the subset sum problem. We harness distributional approximations to model the number of subsets which sum to a certain size. These distributional approximations are formulated in two ways: using bounds to justify normal approximation, and approximating the empirical distribution via density estimation. These approximations can be computed in $O(n)$ complexity, and can increase in accuracy with the size of the input data making it useful for large-scale combinatorial problems. Code is available at \url{https://github.com/KristofPusztai/PerfectSum}.
\paragraph{keywords:} Subset Sum Problem, Perfect Sum Problem, Probability Theory, Central Limit Theorem, Combinatorics, Finite Sample Estimation.
\end{abstract}

\newpage

\section{Introduction}

Combinatorial problems have a notorious reputation for being computationally hard to solve efficiently and, as a result, understanding the underlying properties proves useful in many real life scenarios. The difficulty of solving such problems has made them particularly relevant in the field of cryptography with ciphers such as the well known ENIGMA machine used by Germans during World War II being a prime example\citep{enigma}. Such encryptions make use of of the immense solution space associated with such combinatorial problems, making cracking them a complex task\citep{encryption}. 

Additionally, we can model lots of natural events via a combinatorial approach. Examples where this is especially relevant can be found in combinatorial biology, specifically, the study of gene re-combination, where it is important to understand the dynamics of different possible combinations of gene alterations\citep{combi_bio}\citep{combinatorialgenes}. Phenotypic convergence refers to multiple different combinations of genomic events leading to a similar biological outcome, or trait\citep{phenotype}. How many different ways (i.e. through what combinations of individual genomic events) a particular phenotype may emerge is of interest in disease genetics. Ultimately, any task which involves finding subgroups of choices out of a set of total possible choices will be well modelled by problems such as the one we are examining.

Many past works have focused on finding the most efficient solutions to such problems or investigating specific algorithm behavior. Despite many sophisticated approaches to this problem, the fastest deterministic solution runs in $O(2^{0.3113n})$ complexity\citep{hadash}. There exists a faster algorithm that implements some form of randomization and so inherently relies on probabilistic aspects, but this work does not focus heavily on the topic of probability and instead is interested in the underlying dynamics of their devised algorithm\citep{cao}. In this work, we will focus instead on exploring the purely probabilistic aspects which will allow us to approximate a solution with relative computational ease rather then trying to find the exact solution.

\section{Past Works}
Recent works in this area are mainly focused on fast deterministic and randomized approaches. In fact, such scenarios are commonly used to introduce dynamic programming, specifically, the Bellman method\citep{bellman}. Other solution methods include exhaustive search and divide and-conquer approaches with different modifications yielding sophisticated and efficient algorithms. One such modification to general exhaustive search by \cite{cao} uses a novel data arrangement to find a solution with a computation complexity of $O(2^n)$ and is the first published algorithm which returns all subsets. As a result, this method finds the exact solution to the perfect sum. Additionally, a probabilistic algorithm is also introduced which implements random permutations and a truncation. However, this algorithm only works well for large $n$ and is designed to answer the \hyperref[sec:ss]{decision problem} which is more tractable than finding the number of subsets.

Some more statistically oriented works also exist but are heavily focused on analyzing specific algorithms. The most notable of these is the work of \cite{datri} which investigates the limiting behavior of variants of the GOLOSONE algorithm. This work obtains exact distributions for the output parameters of the algorithms. While the result is certainly interesting, it does not reveal any direct properties of the problem itself. Our work focuses directly on the problem itself and does not deal with any specific solution algorithm.

\section{Problem Formulation}
First we will define the subset sum problem and then the perfect sum problem to show that the latter is an extension of the former.

\subsection{Subset Sum}\label{sec:ss}

The subset sum decision problem is defined formally as: given a set $S = \{x_i, x_j,..., x_n\}$ and some value $T$, decide if there exists a subset of $S$ that sums up to $T$\citep{ssp}. That is, there exists some subset $S_k \subseteq S$, such that $\sum_{x \in S_k}x = T$

\subsection{Perfect Sum} \label{sec:ps}

The perfect sum can be stated as follows: given a set of values and a target sum, we want to find all subsets which sum to our target value. Formally, this can be defined in the following set of conditions.
\\
\\
Given a set of size n, $S = \{x_i, x_j,..., x_n\}$, and some value $T$ we want to find all such subsets $S_k \subseteq S$ so that $\sum_{x \in S_k} x = T$ for all $k = 1, ..., n$. In this paper, we will focus on finding the \emph{total} number of all such subsets instead of the actual subsets themselves.
\\
\\
Clearly this is an extension of subset sum since the solution to the perfect sum gives us sufficient information to solve the decision problem while also providing additional information.

\section{Probabilistic Analysis}\label{sec:probability}

Using our definitions from section \ref{sec:ps}, we can treat any subset $S_k$ as a sample taken without replacement from the set S. In order to inspect the dynamics of the subset further, we can calculate the probability that any value of the set S, $x \in S$, will be in the subset $S_k$. These calculations are not novel, and their proofs can be found in the appendix for rigor. The novelty will come in the application of these calculations in approximating a solution.

\paragraph{\hyperref[sec:proof1]{Lemma 2}:}\label{sec:l1}
The event that value $x \in S$ is chosen to be in $S_k$ has probability $\mathbbm{P}(x\in S_k) = k/n$ where $k$ denotes the cardinality of the subset $S_k$, $k = |S_k|$, and $n$ denotes the cardinality of the set $S$, $n = |S|$. 

This subtle property is simple yet useful as we can use this to show that the expected average of any subset will be the same as the average of the original set.

\paragraph{\hyperref[sec:proof3]{Theorem 1}:} \label{sec:th1}
The expected value of the sum of any subset of size $k$, $\sum_{x \in S_k}x$, is equal to $k$ times the mean of the total set, $\bar{S}$.

Formally:
$$E[\sum_{x \in S_k} x] = k\bar{S}$$

We are now faced with a more difficult task which is to calculate the variance of the sums of some subset of size k. This is made complex by the fact that we are sampling without replacement which induces non-zero covariance among each member of the sample. 

From the formula for variance we have:
$$Var[\sum_{x \in S_k} x] = Var[\sum_{x \in S_k} x] = \sum_{x \in S_k} Var[x] + \sum_{x_i, x_j \in S_k, x_i \neq x_j} Cov[x_i, x_j]$$

Hence, we know that the variance of the sum of random variables $x \in S_k$ will include covariance terms due to the lack of replacement. Intuitively, we know that this covariance should be negative since if we have two chosen values from $S_k$, call them $x_i$ and $x_j$ then we know that if $x_i$ has a large value we can no longer pick this large value when choosing $x_j$. Additionally, we know that the total size of the set we are choosing from should also be present since as the set size approaches infinity, this limit approaches sampling with replacement. With these in mind, we can take a look at the formula for covariance and notice a few things.

$$Cov[x_i, x_j] = E[x_i * x_j] - E[x_i] * E[x_j] \textrm{ \ \ \ for $x_i, x_j \in S_k$ and $ x_i \neq x_j$}$$

Firstly, we note that $E[x_i] = E[x_j]$ for any $x_i, x_j \in S_k$ since all values $x \in S$ have the same probability of being in $S_k$ from our calculations in \hyperref[sec:th1]{Theorem 1}, and all values in $x \in S_k$ have the same probability, $\frac{1}{k}$, of being chosen. In fact, we can show that $E[x] = \bar{S}$, $\forall x \in S$ where $\bar{S}$ denotes the mean of the original size $n$ set. A short proof of this is left in the appendix section \ref{sec:proof1}. Now we are left to find the $E[x_i * x_j]$ term which happens to be something quite convenient as we see in below in \hyperref[sec:l2]{Lemma 2}.

\paragraph{\hyperref[sec:proof4]{Lemma 2}:}\label{sec:l2}
$$E[x_i * x_j] = -\frac{\sigma^2}{n-1} + \bar{S}^2$$

We now have all of our terms in covariance accounted for which allows us to conclude \hyperref[sec:l3]{Lemma 3} by simply plugging in the values from our previous derivations.
\paragraph{\hyperref[sec:proof5]{Lemma 3}:}\label{sec:l3}
$$\textrm{for any $x_i, x_j \in S_k$ we have that }Cov[x_i, x_j] = -\frac{\sigma^2}{n-1}$$

We now have all the components to find the variance of the sum of a subset and by simply plugging in our calculated values from the above sections we arrive at \hyperref[sec:th2]{Theorem 2}.

\paragraph{\hyperref[sec:proof6]{Theorem 2}:}\label{sec:th2}
$$Var[\sum_{x \in S_k} x] = k\sigma^2(1 - \frac{(k-1)}{n-1})$$

This conclusion provides us with significant information about the distribution of possible sum values and is vital to our application for approximating a solution. Note that as $k$ approaches $n$, the variance goes to 0 which intuitively makes sense since we know that once we $k=n$ we are just taking the sum of the whole set.

\section{Application to Finding Solutions}
In the following sections, we provide applications for the above probabilistic properties in finding a solution. We will explore various distributional approximation methods and apply them to calculate a concrete value which we compare to the true answer for solvable randomly generated problems. Psuedocode for our general algorithm structure can be found in the \hyperref[sec:psuedocode]{appendix} and can be seen to have $O(n)$ complexity.
\subsection{Exact Solution}\label{sec:exact}
With the previous calculations in mind, we can try and apply these conclusions to finding a solution to the perfect sum problem. From a probabilistic perspective, if we know the exact discrete distribution of the larger set $S$, then we can calculate the exact distribution of each of the subsets $S_k$. As a result, we can find the percentage of subsets of each size $k$ which sum up to a given value. In fact, with this distributional perspective we gain immense flexibility as we can also find the percentage of subsets which sum to greater, or less then a specified value. Once we have the percentage, we can multiply this percentage by the total number of different possible subsets of size $k$, which is just a combination formula computation, and we are left with the exact number of subsets. 

The usefulness of this approach is limited, though, by the fact that we must calculate exact discrete probability density's. For example, let's examine $k$ = 3. Let's denote the sum of 3 random variables $x_1, x_2, x_3$ and $X = \sum S_3 = x_1 + x_2 + x_3$ then, via the convolution formula of probability, we have:
$$\mathbbm{P}(X=x) = \sum_{x_1 \in S}\sum_{x_2 \in \{S\backslash x_1\}} \mathbbm{P}(x_1 = j)*\mathbbm{P}(x_2 = x - x_1 - x_2 |x_1)*\mathbbm{P}(x_3 = x - x_1 - x_2|x_1, x_2)$$

which leaves us once again, with a combinatorial problem. Ultimately, our previous results are not useful in the pursuit of an exact solution. However, they grant us the ability to employ approximations which can be designed to increase in accuracy as $n$, the size of set $S$, gets larger. In fact, as $n$ gets large, the exact solution becomes infeasible to calculate by nature of combinatorics. Using the the formulas above, we can use methods of distributional approximation such as one of the various versions of the Central Limit Theorem which, under their certain regularity conditions, grant increasing accuracy and are significantly faster to calculate with just $O(n)$ complexity.

\subsection{Naive Approximation via Normal Distribution}
\begin{figure}[H]
	\centering
	\begin{subfigure}[JSD of Normal: 0.04613 \newline JSD of Irwin-Hall: 0.03013]
	{\includegraphics[scale=0.3]{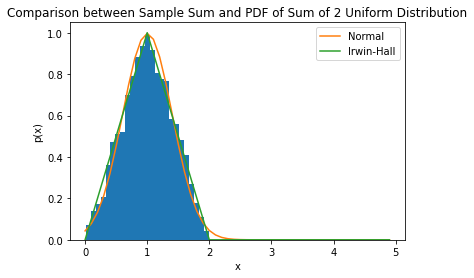}}
	\end{subfigure}
	\begin{subfigure}[JSD of Normal: 0.03118 \newline JSD of Irwin-Hall: 0.02225]
	{\includegraphics[scale=0.3]{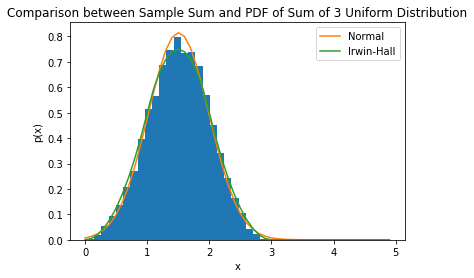}}
	\end{subfigure}\newline
	\begin{subfigure}[JSD of Normal: 0.03164 \newline JSD of Irwin-Hall: 0.02237]
	{\includegraphics[scale=0.3]{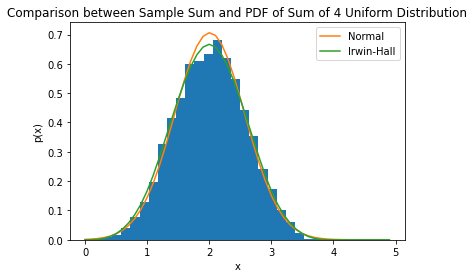}}
	\end{subfigure}
	\begin{subfigure}[JSD of Normal: 0.01849 \newline JSD of Irwin-Hall: 0.01832]
	{\includegraphics[scale=0.3]{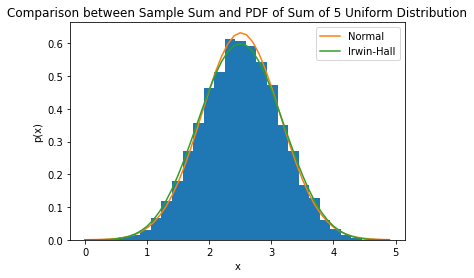}}
	\end{subfigure}
	\caption{In these graphs we see that the normal distribution does well at approximating the distribution of our subset sums, which come from a uniform distribution. Notice, however, that the normal distribution is not such a great approximation for low values of $k$, namely graph (a), $k = 2$ and (b), $k = 3$ show a larger deviation between the normal approximation and the true values. Graph (c), $k = 4$, and (d), $k = 5$, are much better approximated. As the subset size gets larger, the normal pdf will become a more accurate approximation as shown by the Berry-Esseen Bounds\citep{error}.}
\label{fig:fig1}
\end{figure}

A naive approximation approach takes form in a modified version of the Central Limit Theorem (CLT)\citep{clt2}, allowing us to approximate the distribution of the sum of subsets as a normal distribution. This approximation hinges on the assumption that our values in set $S$ are determined by i.i.d random samples taken from some underlying latent distribution. If these assumptions do not hold, we cannot invoke the Berry-Esseen bounds shown below and the accuracy of our approximation will not necessarily increase as set size and threshold value increase. However, if these assumptions do hold, then the accuracy of this approximation will increase the larger $T$ and $S$ are. We can see this empirically in Figure~\ref{fig:fig1}. For smaller $T$, solutions will contain smaller subset groupings and can cause inaccuracy. These cases will depend heavily on the actual distribution of values in S. However, the larger the size of set $S$, $n$, is and the larger our target, $T$, is, the more subsets of larger size $k$, which results in a more accurate approximation via this method as these larger subsets will contain most of the desired sums. More formally, we can theoretically bound the deviation between the distribution we are approximating and the normal distribution. Many of the different proofs of the classical CLT rely on such bounding, examining the absolute difference between the normal distribution and distribution of the sum of i.i.d random variables\citep{clt2}. However, for our case, we are looking at a bound for the sum of i.i.d samples \textit{from a finite population} which we will assume to be a set of independent random variables in this paper. 

With this in mind, we can harness the bound presented in "Berry–Esseen Bound for a Sample Sum from a Finite Set of Independent
Random Variables"\citep{error}:

$$\sup_{x}|\mathbbm{P}(\frac{S_k}{\sqrt{kb}}\leq x)-\Phi(x)| \leq C \min(\delta_1,\delta_2 + 1/\sqrt{kq})$$\

\textit{where $C$ is an absolute constant,\\ $\frac{1}{0} := \infty$, \newline
$p=\frac{k}{n}$,\newline
$q=1-p$,\newline
$b=1-pn^{-1}\sum_{i=1}^n(E(x_i))^2$,\newline
$\delta_1=n^{-1}\sum_{i=1}^n\frac{E(|x_i|^3)}{\sqrt(k)b^{3/2}}$, \newline
$\delta_2=n^{-1}\sum_{i=1}^n\frac{|E(x_i)|^3}{\sqrt{nb}} + n^{-1}\sum_{i=1}^n\frac{E(|x_i - pE(x_i)|^3)}{\sqrt{n}b^{3/2}}$}
\newline
\\
This bound gives us a good idea of the theoretical worst case convergence of our sums to the normal distribution and provides a guarantee that these sums do, in fact, approach the normal distribution as $n$ increases. However, calculating this theoretical deviation does not give us much information on how well the normal approximation is doing with a specific data-set. For a more concrete and specific result, we can use the discrete Jensen-Shannon divergence metric\citep{js} over the desired data which allows us to quantify how well the normal approximation is doing with specified data distributions. The discrete Jensen-Shannon divergence metric over probability space $\Omega$ is defined as:

$$JSD(\mathbbm{P}||\mathbbm{Q}) = \frac{1}{2}\sum_{x\in \Omega}\mathbbm{P}(x)log(\frac{\mathbbm{P}(x)}{M(x)}) + \frac{1}{2}\sum_{x\in \Omega}\mathbbm{Q}(x)log(\frac{\mathbbm{Q}(x)}{M(x)})$$
\begin{center}where $M(x) = \frac{1}{2}(\mathbbm{P}(x) + \mathbbm{Q}(x))$\end{center}

Note that there are many different forms of CLT's all with their own assumptions and convergence rates. In fact, there are also general methods for bounding the difference between distributions. Specifically, Stein's method\citep{stein} provides a way to calculate bounds of the difference between normal distribution and any other distribution we might assume our data comes from. As a result, this naive approximation can still useful and provide accurate enough approximations with minimal computation. In section \hyperref[sec:5.4]{5.4} we will introduce a an approximation which does not hinge on these assumptions and instead tries to estimate the density directly via sampling.

\subsection{Improved Approximation}
For scenarios where our set $S$ can be approximated well via some other continuous distribution besides the normal, we can make our approximation even more accurate. Formally, if the original set is an i.i.d sample of some known continuous distribution, then we can use this to find a better distribution of the sums of each different subset size. This will result in a more accurate approximation, especially for subsets of smaller size $k$, since this is where the CLT approximation is the least accurate.

\begin{figure}[H]
	\centering
	\begin{subfigure}[JSD of Normal: 0.09837 \newline JSD of Chi-Square: 0.02314]
	{\includegraphics[scale=0.35]{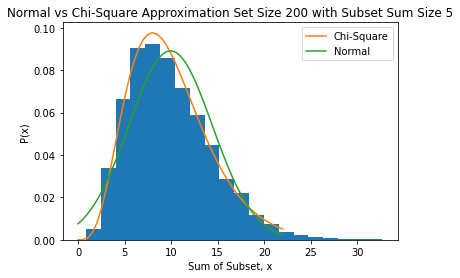}}
	\end{subfigure}
	\begin{subfigure}[JSD of Normal: 0.01370 \newline JSD of Chi-Square: 0.15208]
	{\includegraphics[scale=0.35]{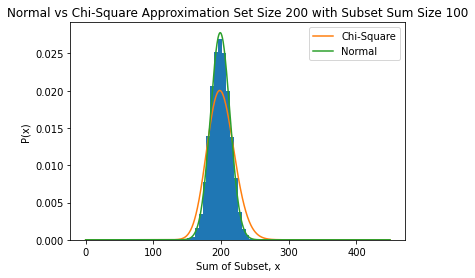}}
	\end{subfigure}\newline
	\begin{subfigure}[JSD of Normal: 0.09587 \newline JSD of Chi-Square: 0.01658]
	{\includegraphics[scale=0.34]{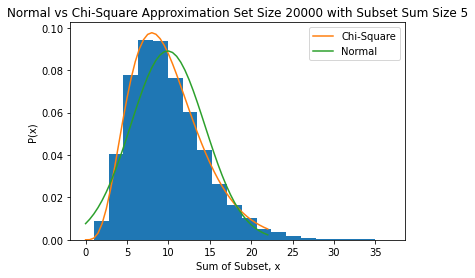}}
	\end{subfigure}
	\begin{subfigure}[JSD of Normal: 0.03234 \newline JSD of Chi-Square: 0.01357]
	{\includegraphics[scale=0.34]{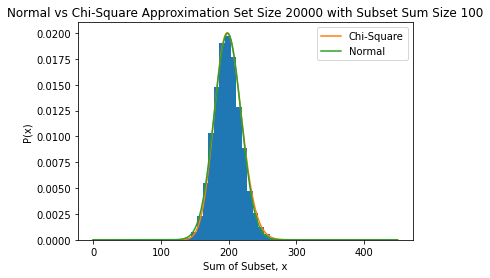}}
	\end{subfigure}
	\caption{Notice in the simulated set of size 200 that there is a slight deviation between the theoretical distribution and sample distribution in picture (a). This causes the approximation to be off for larger subset sums as seen in (b), where the normal approximation has a lower JS-Divergence. In (c) and (d) the set size is much larger, 20,000 samples, which leads to a better approximation and justifies the use of the alternate distribution over the normal.}
	\label{fig:fig6}
\end{figure}

However, we must be careful with this as the normal distribution can still be a better approximation if the size of $S$ is small. We want the variance of this sample mean to not be small to ensure that this alternate distribution is truly better as even slight deviations in sample values can cause this alternate distribution to fail at accurately describing larger subset sums. This is due to the fact that smaller sets will have higher variance in means. We can see this directly through a simple examination of the formula for calculating sample means of i.i.d random variables:
$$\bar{X} = \frac{1}{n}\sum_{i=1}^n x_i \rightarrow Var(\bar{X}) = \frac{Var(x_i)}{n}$$
However, since the normal approximation takes the mean and standard deviation of the samples themselves into account, it is able to better correct for these deviations when $S$ is small. We can clearly see this in Figure~\ref{fig:fig6} which demonstrates the effects of set size on accuracy of using the underlying distribution from which it was sampled.
To quantify when using a distribution other then the normal would perform better, we can again take an empirical approach and use the discrete Jensen-Shannon divergence metric\citep{js} to measure how well the two distributions approximate our subsets. This will then allow us to make exact comparisons between the two distributions approximation abilities for our specific data. Additionally, we can take a theoretical approach and note that we want the mean of our sample set $S$ to be close to the actual mean of the distribution. Hence, we want some some size $n$ sample which results in a low variance of $\bar{X}$. The value of $n$ will completely depend on the desired threshold, and the variance of the underlying distribution from which samples are drawn.

\subsection{Empirical Approximation Via Density Estimation}{\label{sec:5.4}}
For situations where any form of CLT may not be applicable and normal distribution does not provide a good approximation of the densities, we can estimate densities directly through sampling and finding an empirical distribution. There are several methods both parametric and non-parametric for density approximation. Parametric approaches include fitting a Pearson distribution\citep{pearson}, Johnson-$S_u$ distribution\citep{johnson} or a variety of distributions which can be fitted via either bayesian or frequentist methods. Non-parametric approaches include the well known method of kernel density estimation, which is what we will employ in our implementation due to its flexible nature. Specifically, we will use a tophat kernel with bandwidth, h, specified by the 10\% quantile of the difference between data points. The estimated probability for some threshold value $t$ is then calculated as follows:

$$\mathbbm{P}(X = t) = \frac{1}{nh}\sum_{i}^{n}K(\frac{t-x_i}{h})$$

where $K(x)$ signifies the kernel function with the appropriate properties of \(\int_{-\infty}^{\infty} K(x) \,dx\) = 1 and symmetry, $K(-x) = K(x)$.

We use this to estimate the probability density based on sampling without replacement from the set $S$ and taking the sums. As a result, we are directly estimating the distribution of the sum. The choice of kernel density estimation is due to the flexibility that this estimation method provides, as it makes no assumptions about the data unlike some other parametric techniques. However, this flexibility comes at a cost as the number of samples, choice of bandwidth and choice of kernel will heavily affect the shape and accuracy of the estimation. Finding the optimal parameters will likely require extensive exploratory data analysis. Additionally, calculating such kernel density estimations depends on the number of data points and more data increases run-time.

\section{Simulations}
 There are three cases to test our approximation accuracy on generated sets. The first case is where the naive approximation does well. We explore both continuous and discrete cases. We find that in discrete case an improved approximation method may not exist and that our naive method performs the best. Specifically, we simulated sets from a discrete scaled uniform distribution, $U(0,20)$, to create our initial set. In the next section, \hyperref[ss:naive]{(6.1)}, we show that the naive normal approximation here provides better results then by trying to model this discrete distribution by its continuous counterpart.

However, for cases in which $S$ is drawn from a continuous underlying distribution, we can use this continuous distribution which is a better approximation than the normal. In \hyperref[ss:improved]{Section 6.2} we will explore the case where $S$ is drawn from chi-square distribution which is known to be heavily skewed and so will render our naive normal approximation less accurate. From our simulations, we can see where our approximations work well, and where they don't.

Lastly, in \hyperref[ss:empirical]{Section 6.3}, we will explore cases where independence does not hold and observe how these effect our normal approximation and kernel density approximations.

\subsection{Cases for Naive Approximation}\label{ss:naive}
\begin{figure}[H]
	\centering
	\begin{subfigure}[JSD of Normal: 0.12712 \newline JSD of Irwin-Hall: 0.12986]
	{\includegraphics[scale=0.22]{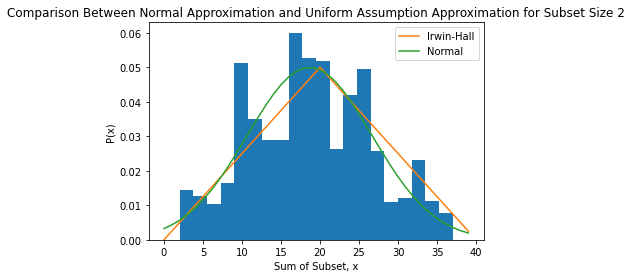}}
	\end{subfigure}
	\begin{subfigure}[JSD of Normal: 0.08121 \newline JSD of Irwin-Hall: 0.10295]
	{\includegraphics[scale=0.22]{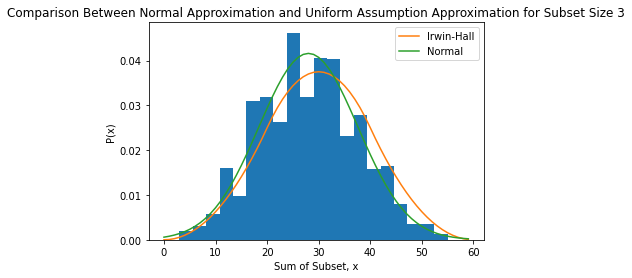}}
	\end{subfigure}\newline
	\begin{subfigure}[JSD of Normal: 0.05145 \newline JSD of Irwin-Hall: 0.09945]
	{\includegraphics[scale=0.22]{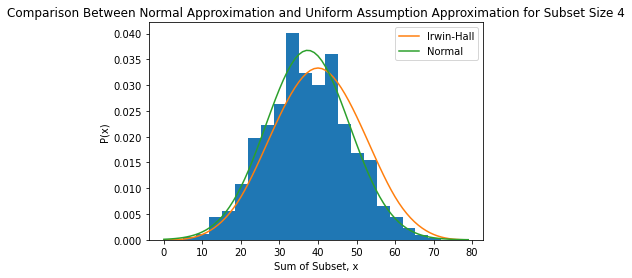}}
	\end{subfigure}
	\begin{subfigure}[JSD of Normal: 0.02421 \newline JSD of Irwin-Hall: 0.10691]
	{\includegraphics[scale=0.22]{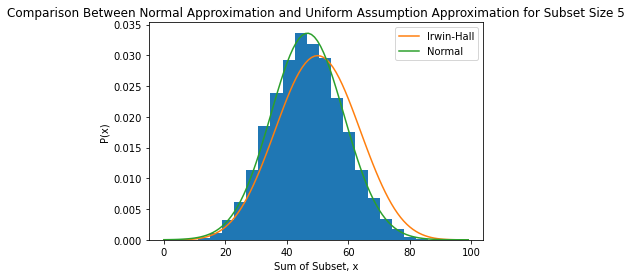}}
	\end{subfigure}
	\caption{The Normal approximation works better for our discrete example since it directly takes into account mean and standard deviation of the discrete set. This discrete set was drawn from a discrete uniform distribution and one would think trying to approximate via a continuous uniform distribution would yield relatively good results. However, the Irwin-Hall distribution (sum of uniform distributions) does not do a good job at approximation and cannot account for certain properties resulting from the discreteness of the data.}
\label{fig:fig4}
\end{figure}

For cases where the values of set $S$ are constrained to discrete values, it may not be feasible to find a more accurate continuous approximation then the normal distribution. A demonstration of this can be seen in Figure~\ref{fig:fig4} in the case of the discrete uniform distribution. Additionally, if no information about the underlying distribution is known, then the Naive approximation is certainly one approach to modelling the distribution of sums and, theoretically, will be extremely effective for large sets and a large target threshold. 

In our example, we use a scaled discrete uniform distribution, $U(0,20)$, and investigate the approximation abilities of the normal distribution. One would think initially that using a continuous uniform distribution to approximate this would provide good results, but we must be careful with this. For example, let's consider a uniform (0,1) discrete distribution. Note that 0 and  1 both have probability of 0.5 of being picked. Then if we sum two independent samples from this distribution, the probability of the sum being 2 will be 0.25. If we tried to approximate this via a continuous uniform distribution, we would estimate the probability of the sum being 2 to be 0, hence not a very useful approximation. 

\begin{figure}[H]
    \centering
    \includegraphics[scale=0.45]{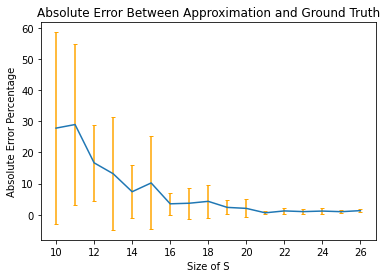}
    \caption{This graph plots the average absolute error percentage between ground truth and approximation values for 20 set simulations. Notice a decreasing trend in absolute approximation error as set size increases. This is due to the fact that subsets with larger size begin to dominate the solution and allow for better approximation. As an increasing number of solutions are found within larger subsets, our approximation becomes more accurate. Orange deviation lines signify 1 SD for 20 different simulations.}
    \label{fig:fig3}
\end{figure}

Due to computational constraints, the ground truth solution for the perfect sum problem can only be calculated up to set $S$ sizes of 26. However, even with such a small set size, we see some trends emerge. Specifically, in Figure~\ref{fig:fig3} we see that as the set size increases, our absolute approximation error decreases.
\newpage

\subsection{Cases for Improved approximation}\label{ss:improved}

For cases where the values of set $S$ can take on continuous values, we may find a continuous distribution which approximates these values better then the normal distribution, but this only works well for a significantly large n. This is especially the case if the distribution is highly skewed. Another aspect to consider when approaching the continuous case is that finding the number of exact sums results in an exact solution of 0. In fact, We get more insight from looking at how many sets sum up to greater/less then a certain threshold value. We performed various simulations which explore the validity of our approximation approach and found promising results. This approximation does well when the threshold is higher up relative to the set size, since larger subsets will contribute more to the solution.
\begin{figure}[H]
	\centering
	\begin{subfigure}{}
    	\includegraphics[scale=0.4]{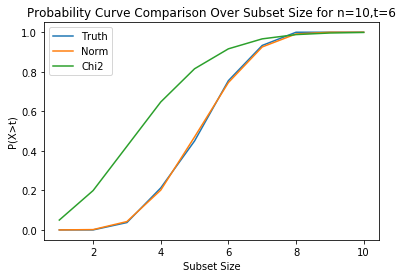}
	\end{subfigure}
	\begin{subfigure}{}
    	\includegraphics[scale=0.4]{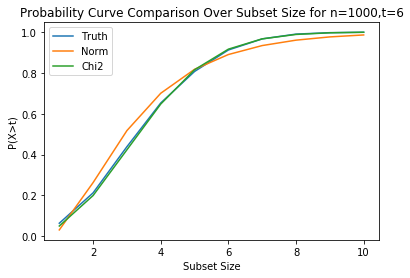}
	\end{subfigure}
	\begin{subfigure}{}
    	\includegraphics[scale=0.4]{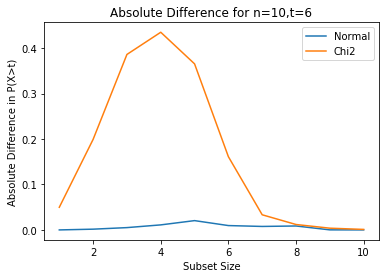}
	\end{subfigure}
	\begin{subfigure}{}
    	\includegraphics[scale=0.4]{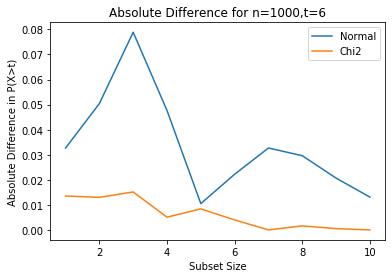}
	\end{subfigure}{}
	\caption{The Normal approximation works better for smaller $n$ due to small sample variation which the chi-squared approximation is not able to adjust for despite the samples coming from an underlying chi-square distribution. However, as $n$ increases, the set follows a chi-squared distribution more closely and the roles reverse, with the chi-squared approximation becoming optimal.}
	\label{fig:chivnormal}
\end{figure}

Additionally, the normal approximation did better on smaller input set sizes when compared to the chi-squared approximation. This is likely due to the fact that the normal approximation directly takes the mean and standard deviation of the set as parameters making it highly flexible when faced with sample deviations which occur in small samples sizes. However, as the set size increased, the chi-squared approximation performed better, with lower approximation error. This is depicted in Figure~\ref{fig:chivnormal}. Note, however, that despite the normal approximation becoming less accurate than the chi-squared for larger n, it is still a relatively good approximation, with an maximum absolute error of ~0.08, while the Chi-Squared approximation was off by 0.4 in the worst case. The normal approximation is versatile as it's parameters allow for sample deviation adjustments.

\subsection{Cases for Empirical Approach}\label{ss:empirical}
\begin{figure}[H]
\centering
	\begin{subfigure}[JSD of Normal: 0.79811 \newline JSD of KDE: 0.27923]
	{\includegraphics[scale=0.32]{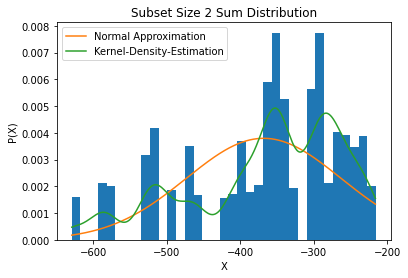}}
	\end{subfigure}
 \hfill
	\begin{subfigure}[JSD of Normal: 0.82409 \newline JSD of KDE: 0.10890]
	{\includegraphics[scale=0.32]{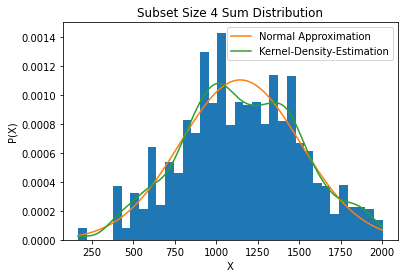}}
	\end{subfigure}
 \hfill
	\begin{subfigure}[JSD of Normal: 0.79985 \newline JSD of KDE: 0.13563]
	{\includegraphics[scale=0.32]{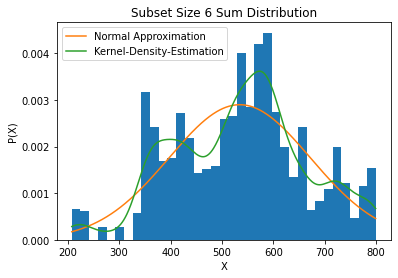}}
	\end{subfigure}
 \hfill
	\begin{subfigure}[JSD of Normal: 0.78635 \newline JSD of KDE: 0.23210]
	{\includegraphics[scale=0.32]{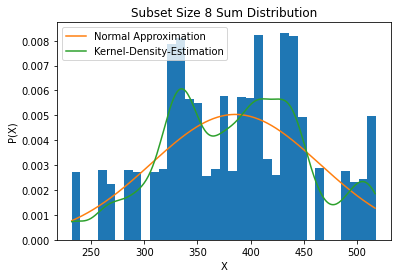}}
	\end{subfigure}
	\caption{Note the KDE's ability to account for large gaps between the possible sums such as cases like (a) and (d). The KDE provides a better estimation of such distributions then the normal and achieves a lower JS-Divergence.}
	\label{fig:kde}
\end{figure}

For cases where no form of the CLT can be invoked and we have no a priori knowledge of the distribution, we are faced with two options: continue using the normal, or estimate the distribution empirically. There are several benefits to using the normal approximation including that it will not depend on the accuracy/validity of taking a sample to estimate the empirical distribution. As a result, no randomness will be involved and this approach will always give back the same answer for a given set of inputs. Additionally, the existence of bounds on the normal approximation error for dependent samples has been shown in Stein\citep{stein}. However, if no bounds can be shown and the normal does not seem to be a good approximation then an empirical approach might best suit the situation. Specifically, in situations where the data follows complex distributions such as the one seen in Figure~\ref{fig:kde}, our empirical method significantly outperforms the naive normal approximation.
\section{Applications To Other Problems}
This solution approach can be applied to any problem which can be boiled down to a perfect sum-type question. These questions include the classic "knapsack" and "change-making" problems\citep{changemaking}\citep{knapsack}. However, we must bear in mind that, as mentioned in the problem statements section, our approach is focused on determining the \emph{total} number of possible solutions, and not actually finding these solutions themselves. This leads to the major drawback of this method in that, while we may have an accurate approximation of this total value, we are not any closer to knowing what these sets actually are. In fact, there are no truly efficient(sub-exponential time) algorithms to find these exact solution sets and so in big-data settings, finding all possible sets becomes infeasible, not only computation-wise, but also memory wise as the number of solutions can grow to hundreds of millions very rapidly(i.e. 100 choose 5 =  75,287,520 different possible combinations). With this in mind, one can apply our approach to help narrow down the computational scale, focusing on desired subset-sizes which are likely to contain solutions. For example, to find the minimum subset size which sums to a desired number, our approach can be used to see whether a subset of size $k$ is likely to sum up to this desired value. We can find the minimum size of a set where the predicted number of sets summing up to our target value is 1 via our estimation method. We can then pair this information up with an exact solution finding algorithm that actually examines the sets to find a solution and focus on on just subsets of the sizes we've identified from the previous step. In this way, we've reduced the computation time significantly as we will only need to focus on a limited range of different subset sizes instead of all possible sizes.

Examining real world applications of this approach, it is immediate that this is particularly relevant to the field of genomics research due to the combinatorial nature of gene inheritance/recombination. There is a whole sub-field of biology which focuses on combinatorial methods for gene interaction\citep{biology}. Research such as "Combinatorial control of gene expression" demonstrates the relevance of such ideas in understanding genomics\citep{combinatorialgenes}. Specifically, the ideas presented in this paper are useful if one wants to find the number of different possible genes which could result in a specific event occurring. We need only to assign genes a numerical value in their contribution to this event, and then we can apply our approach to find the total number/percentage of potentially threshold surpassing combinations. This could be especially useful for cancer research, where we can think about each gene combination as contributing a certain amount, and once a stability level is crossed, a cell turns cancerous. These applications I will leave for further thought and exploration to the reader.
\newpage

\section{Conclusion}

In this paper, we introduce a novel approximation approach to the perfect sum problem which is focused on finding the total number of sub-sets which sum to/surpass a desired threshold. This approach harnesses probability theory to estimate the distribution of the sums of subsets. Specifically, we examine the use of the normal distribution as well as non-parametric density estimation to estimate the distribution of the sums of different sized subsets when no a priori distribution information is known. Our algorithm, presented in appendix section \hyperref[sec:psuedocode]{\textbf{Perfect Sum Psuedocode}} runs in $O(n)$ complexity and can increase in accuracy as set size increases if certain regularity conditions hold. As a result, it directly addresses the need for good approximation in big data cases where finding the exact solution to combinatorial problems is infeasible.

\section{Appendix}

\subsection{Proof Of Expected Value For Members of Subset Size k}\label{sec:proof1}
for any $x \in S_K$ we have: $$E[x] = \sum_{x \in S} x * \mathbbm{P}(x)$$ 
Note that $\mathbbm{P}(x)$ denotes the probability of picking the value x from all \textit{possible} values in subset $S_k$ which is the joint event of this value ending up in $S_k$ first and then being picked out of all other k-1 values
$$= \sum_{x \in S} x * 1/k * k/n$$ $$= \sum_{x \in S} x * 1/n$$ $$= 1/n \sum_{x \in S} x = \bar{S}$$
\qedsymbol
\subsection{Proof of Lemma 1}\label{sec:proof2}

Notice that in a subset of size $k$, for some value $x \in S$ there are $n-1 \choose k-1$ different possible subsets $S_k$ which satisfy $x \in S_k$. There are a total of $n \choose k$ subsets of size k. Thus, the probability of choosing a subset which contains $x$ out of all possible subsets of size $k$ is:

$$\mathbbm{P}(x \in S_k) = \frac{{n-1 \choose k-1}}{{n \choose k}} = \frac{\frac{(n-1)!}{(k-1)!(n-k)!})}{\frac{n!}{k!(n-k)!}} = \frac{k!(n-1)!}{(k-1)!n!} = k/n$$

\qedsymbol
\subsection{Proof of Theorem 1}\label{sec:proof3}

$$\sum_{x \in S_k} x = \sum_{x \in S_k} x = \sum_{x \in S} x * \mathbbm{1}\{x \in S_k\} \textrm{ \ \ \ where } \mathbbm{1}\{x \in S_k\} =
  \left\{ \begin{array}{l}
    1 \textrm{ if }x \in S_k \\
    0 \textrm{ otherwise}
  \end{array}\right.
$$

Then we have that:

$$E[\sum_{x \in S_k} x] = E[\sum_{x \in S} x * \mathbbm{1}\{x \in S_k\}]$$ $$= \sum_{x \in S} x * E[\mathbbm{1}\{x \in S_k\}]$$ $$= \sum_{x \in S} x * \mathbbm{P}(x_i \in S_k)$$ $$= \sum_{x \in S} x * k/n$$ $$= \frac{k}{n}\sum_{x \in S} x = k\bar{S} $$

\qedsymbol
\subsection{Proof of Lemma 2}\label{sec:proof4}
$$E[x_i * x_j] = \sum_{x_i, x_j \in S, x_i \neq x_j} x_i * x_j * \mathbbm{P}(x_i, x_j \in S_k) * \mathbbm{P}(x_i, x_j)$$
Following the notation from \hyperref[sec:l1]{Lemma 1}, $\mathbbm{P}(x_i, x_j \in S_k)$ denotes the probability of the event that values $x_i$ and $x_j$ get chosen to be in the subset $S_k$. $\mathbbm{P}(x_i, x_j)$ denotes the probability of the event that values $x_i$ and $x_j$ get chosen from the values in the subset $S_k$. We need these two events because in order for the ultimate event of $x_i$ and $x_j$ getting chosen from $S_k$, we need to ensure that they actually are in $S_k$ to begin with, otherwise they cannot be chosen so we need to capture the joint event. Expanding on these probabilities further:
$$\mathbbm{P}(x_i, x_j \in S_k) = \frac{k}{n} * \frac{k-1}{n-1}$$
$$\mathbbm{P}(x_i, x_j) = \frac{1}{k(k-1)}$$
Hence,
$$\sum_{x_i, x_j \in S, x_i \neq x_j} x_i * x_j * \mathbbm{P}(x_i, x_j \in S_k) * \mathbbm{P}(x_i, x_j) = \sum_{x_i, x_j \in S, x_i \neq x_j} x_i * x_j * \frac{1}{n(n-1)}$$
$$ = \frac{1}{n(n-1)}(\sum_{x_i \in S} x_i * \sum_{x_j \in S, x_j \neq x_i} x_j)$$
$$= \frac{1}{n(n-1)}(\sum_{x_i \in S} x_i * (\sum_{x_j \in S} x_j - x_i))$$
$$= \frac{1}{n(n-1)}(\sum_{x_i \in S} x_i * (\sum_{x_j \in S} x_j - x_i))$$
$$= \frac{1}{n(n-1)}(\sum_{x_i \in S} x_i * \sum_{x_j \in S} x_j - \sum_{x_i \in S} x_i^2) \textrm{ \ \ \ note that \ $\sum_{x_i \in S} x_i * \sum_{x_j \in S} x_j = nE[x_i]*nE[x_j]$}$$
$$= \frac{1}{(n-1)}(n^2E[x_i]E[x_j] - E[x_i^2])$$
$$=\frac{1}{(n-1)}(n\bar{S}^2 - \sigma^2 - \bar{S}^2)$$
$$=\frac{n}{(n-1)}\bar{S}^2 - \frac{1}{(n-1)}\sigma^2 - \frac{1}{(n-1)}\bar{S}^2$$
$$=\bar{S}^2 - \frac{\sigma^2}{(n-1)}$$
\qedsymbol
\subsection{Proof of Lemma 3}\label{sec:proof5}

Plugging everything into our formula for covariance we have:

$$Cov[x_i, x_j] = E[x_i * x_j] - E[x_i] * E[x_j] \textrm{ \ \ \ for $x_i, x_j \in S_k$ and $ x_i \neq x_j$}$$
$$ =-\frac{\sigma^2}{n-1} + \bar{S}^2 - \bar{S}^2$$
$$=-\frac{\sigma^2}{n-1}$$
\qedsymbol
\subsection{Proof of Theorem 2}\label{sec:proof6}

$$Var[\sum_{x \in S_k} x] = \sum_{x_i, x_j \in S_k} Cov[x_i, x_j]$$ $$= \sum_{x \in S_k} Var[x] + \sum_{x_i, x_j \in S_k, x_i \neq x_j} Cov[x_i, x_j] $$ $$= k\sigma^2 + k(k-1)\frac{-\sigma^2}{n-1} = k\sigma^2(1 - \frac{(k-1)}{n-1})$$
\qedsymbol

\subsection{Perfect Sum Psuedocode}\label{sec:psuedocode}
\begin{algorithm}[H]
	\caption{Perfect Sum Probabilistic Approximation} 
	\begin{algorithmic}[1]
	    \State Compute relevant starting values such as mean, $\mu$, and variance, $\sigma^2$, of empirical distribution of $S$
		\\
		\State Create variable, $R = 0$, to keep track of total number of subsets which satisfy sum conditions regarding $T$
		\For {$k=1,2,\ldots \textrm{length}(S)$}
			\State Compute values (i.e. $\hat{\sigma^2}$, $\hat{\mu}$ via \hyperref[sec:th1]{Theorem 1}, \hyperref[sec:th2]{Theorem 2}) necessary for distribution
			
			approximation choice
			\\
			\State Compute distribution approximation of sum of subsets size $k$
			\\
			\State Use distribution approximation to find desired probability, 
			
			$P = \mathbbm{P}(\sum_{x \in S_k} x  \{=,\geq,\leq\} T)$, of given sum threshold, $T$
			\\
			\State Convert probability to number of subsets, $s = integer(P * {n \choose k})$
			\\
			\State Add to variable $R$ to keep track of total over all different sized subsets, $R = R + s$
		\EndFor
	\State return R
	\end{algorithmic} 
\end{algorithm}
\subsection{Figure 1}
Comparisons were generated by first simulating a set of 500 samples from a uniform distribution. We then sampled 10,000 sums of corresponding sized samples from this set (sums of sample sizes 2,3,4 and 5) and use the properties we found in section \ref{sec:probability} to find the mean and variance for the normal approximation. We use the Jensen-Shannon Divergence (JSD) as the metric for approximating the distribution of sums and compared the resulting normal approximation JSD to that of an Irwin-Hall distributional approximation.
\subsection{Figure 2}
Comparisons were generated by first simulating a 2 sets of samples from a Chi-Squared distribution, one of size 200 and the other of size 20,000. We then sampled 20,000 sums of corresponding sized samples from this set (sums of sample sizes 5 and 100) and use the properties we found in section \ref{sec:probability} to find the mean and variance for the normal approximation. We use the Jensen-Shannon Divergence (JSD) as the metric for approximating the distribution of sums and compared the resulting normal approximation JSD to that of a Chi-Square approximation.
\subsection{Figure 3}
Comparisons were generated by first simulating a set of 27 samples from a uniform distribution scaled to be between 0 and 20. We then sampled 20,000 sums of corresponding sized samples from this set (sums of sample sizes 2,3,4 and 5) and use the properties we found in section \ref{sec:probability} to find the mean and variance for the normal approximation. We use the Jensen-Shannon Divergence (JSD) as the metric for approximating the distribution of sums and compared the resulting normal approximation JSD to that of an Irwin-Hall distributional approximation.
\subsection{Figure 4}
We simulated 20 sets each of sizes 10 all the way to 27 and scaled the desired target value of the subset sum by the size of the set, that is, $T$, was increased as the size of the simulated set was increased. This was done to show that subset sums with larger sizes are approximated more accurately. Specifically, $T = \frac{n*60}{27}$ where $n$ represents the size of the set. We then use our proposed algorithm specified above in section \ref{sec:psuedocode} to calculate the approximated number of subset sums, and compare it to a simple, greedy algorithm which finds the ground truth number of subset sums satisfying $T$.
\subsection{Figure 5}
We compared several different probabilistic characteristics of our approximations, including the empirical CDF of our normal approximation, as well as the absolute difference between the mean of the ground truth distribution and our different approximation methods. We simulated 2 sets with 1,000 and 100 i.i.d samples drawn respectively from a Chi-Square distribution with 2 degrees of freedom. Then, we sampled 100 subset sums of size 1 to 11, recording the ratio of sums which were above a threshold value of 6 to get an approximate ground truth distribution of the ratio of sums which were above 6 in a random sample of size 100. Finally, we compared the absolute difference between the mean of this empirical ground truth distribution and that of our approximation methods, namely, using a Chi-Square distribution vs. a normal distribution for approximation. We see that for small set sizes, the normal distribution is more appropriate for approximation, and as the set size increases, the Chi-Square becomes more suitable.
\subsection{Figure 6}
We start with a set of 10 integer values sampled from a normal distribution whose mean is the previous sample value, and a standard deviation of 100. In this way, we have a complex underlying distribution which is not easily parameterized and perfect for showcasing the benefit of empirical estimation. We then randomly sample sums of sizes 2, 4, 6 and 8 5,000 times each to get an accurate empirical distribution of the subset sums. Finally, we can compare the Jensen-Shannon Divergence (JSD) between the normal approximation and a kernel density approximation for each case to show the kernel density estimation method's superior performance.
\section*{Acknowledgement}
The author thanks Drs. L. Pusztai, S. Mukherjee, M. Marczyk, and C. Hatzis for discussing this problem and providing feedback.

\bibliographystyle{apalike}

\bibliography{biblio}

\end{document}